\documentclass[12pt,twocolumn]{emulateapj}

\newcommand{\beq}{\begin{equation}}
\newcommand{\eeq}{\end{equation}}

\newcommand{\hi}{H{\sc i}~}

\newcommand{\citei}[1]{\citeauthor{#1} \citeyear{#1}}

\usepackage{color}
\usepackage[usenames,dvipsnames]{xcolor}
\usepackage[colorlinks,urlcolor=blue,citecolor=blue,linkcolor=blue]{hyperref}
\usepackage{amsmath}
\slugcomment{}

\begin{document}
 
\title{Dust in the Circumgalactic Medium of Low-Redshift Galaxies}
\date{\today}

\author{
J.E.G. Peek\altaffilmark{1,2}
Brice M\'enard\altaffilmark{3,4,5}
Lia Corrales\altaffilmark{6}
}
\altaffiltext{1}{Space Telescope Science Institute, 3700 San Martin Dr, Baltimore, MD 21218, USA, jegpeek@stsci.edu}
\altaffiltext{2}{Department of Astronomy, Columbia University, New York, NY, USA}
\altaffiltext{3}{Department of Physics \& Astronomy, Johns Hopkins University, 3400 N. Charles Street, Baltimore, MD 21218, USA}
\altaffiltext{4}{Institute for the Physics and Mathematics of the Universe, Tokyo University, Kashiwa 277-8583, Japan}
\altaffiltext{5}{Alfred P. Sloan Fellow}
\altaffiltext{6}{Massachusetts Institute of Technology Kavli Institute for Astrophysics and Space Research, 77 Mass Ave, 37-241 Cambridge, MA 02139, USA}

\begin{abstract}
Using spectroscopically selected galaxies from the Sloan Digital Sky Survey we present a detection of reddening due to dust in the circumgalactic medium of galaxies. We detect the mean change in the colors of ``standard crayons" correlated with the presence of foreground galaxies at $z\sim0.05$ as a function of angular separation. Following \cite{2010ApJ...719..415P}, we create standard crayons using passively evolving galaxies corrected for Milky Way reddening and color-redshift trends, leading to a sample with as little as 2\% scatter in color. We devise methods to ameliorate possible systematic effects related to the estimation of colors, and we find an excess reddening induced by foreground galaxies at a level ranging from 10 to 0.5 millimagnitudes on scales ranging from 30 kpc to 1 Mpc. We attribute this effect to a large-scale distribution of dust around galaxies similar to the findings of \cite{menard10}. We find that circumgalactic reddening is a weak function of stellar mass over the range $6 \times 10^9 M_\odot$ -- $6 \times 10^{10} M_\odot$ and note that this behavior appears to be consistent with recent results on the distribution of metals in the gas phase.
\end{abstract}

\keywords{are useless}

\section{Introduction}\label{intro}

\begin{figure*}[ht]
\includegraphics[scale=1.0]{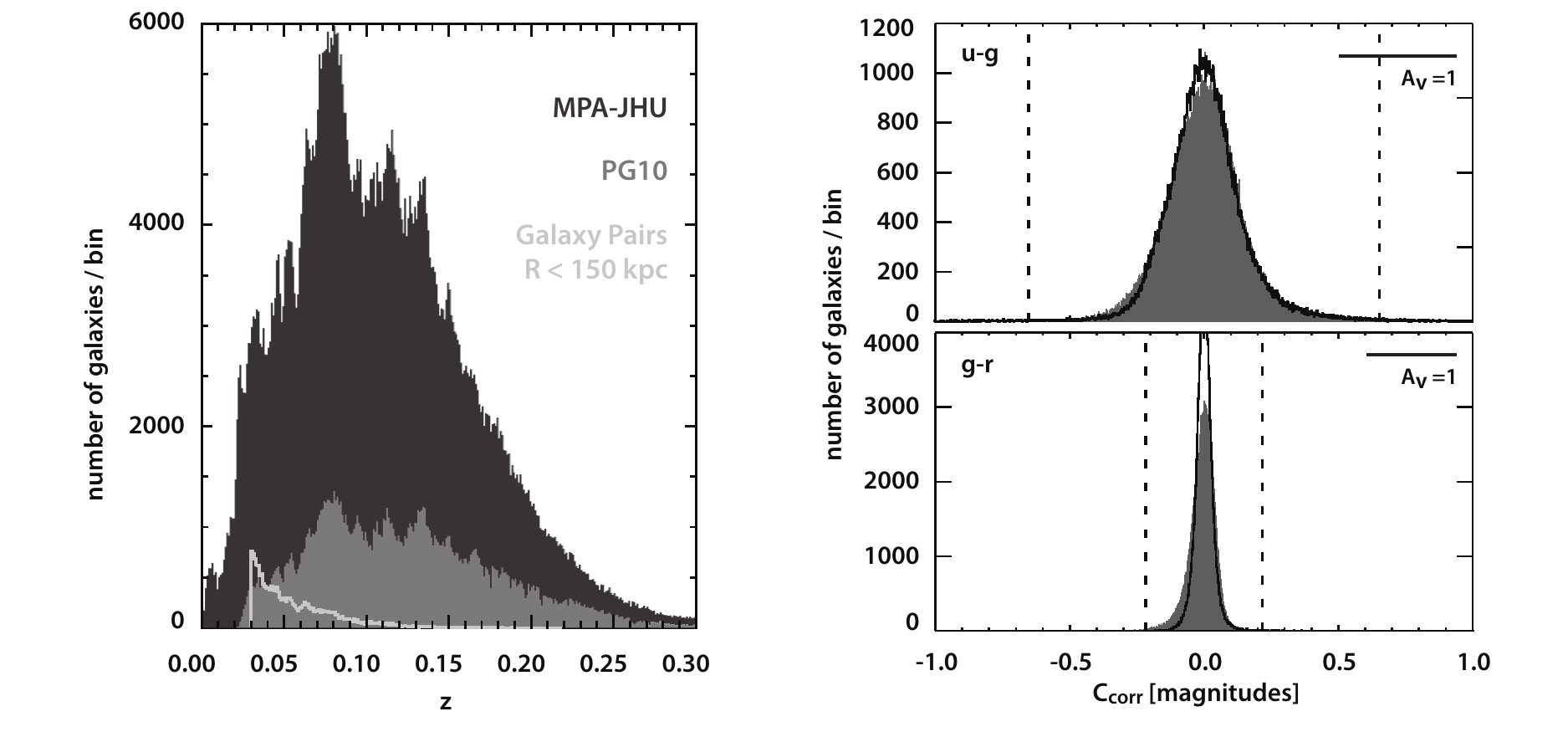}
\caption{\emph{Left:} redshift distribution of the selected galaxies. The dark histogram represents all the foreground galaxies from the Main Galaxy Sample of SDSS DR7. The quiescent galaxies from PG10 are shown in gray. The lightest colored unfilled histogram shows a distribution of the foreground galaxies that have at least one background galaxy within 150 kpc impact parameter. \emph{Right:} the color distributions of PG10 galaxies. Histograms of original PG10 colors (black), and $C_{\rm corr}$, (Equation \ref{cprimeab}; gray, filled), are shown for $u-g$ (top) and $g-r$ (bottom) colors. The expected reddening for Milky Way dust with an $A_v$ of 1, $R_v$ = 3.1 is shown by the black bar. The truncation used to excise extreme outliers is shown in vertical dashed lines.}
\label{galz_cdisp}
\end{figure*}

Galaxies process and return a significant fraction of their accreted gas to their surroundings, the circumgalactic medium (CGM) but the physical mechanisms involved as well as the matter distribution in this environment are still poorly constrained. Much of our knowledge of the distribution of baryons in 

the CGM comes from absorption line studies which probe the gas phase. The interpretation of such measurements is often limited by our lack of knowledge of the ionization state of the gas. In this work we explore the distribution of dust around galaxies and use it as an alternative tracer of metals in this environment, independent of ionization corrections. It is important to realize that a substantial fraction of CGM metals might be in the solid phase. In the ISM, about 30\% of the metals are found in dust \citep{2001ApJ...548..296W}. Measuring dust reddening effectsfrom the CGM may also allow us to put constraints on the grain size distribution and potentially the mechanisms responsible for their ejection from galactic disks to halos, for example supernova explosions \citep{Silk1997,Efs2000} or radiation pressure \citep[e.g.][]{MQT2005,Salem2013}.

A number of authors have reported the presence of dust well beyond galaxy disks. Using superpositions of foreground/background galaxies, \cite{2009AJ....137.3000H} detected dust extinction up to about five times the optical extent of spiral galaxies. Using deep \emph{Herschel} observations, \cite{Roussel:2010dv} showed that emission from cold dust is seen up to 20 kpc from the center of M82. Using UV light scattered by dust grains, \cite{HodgesKluck:2014vn} reported the detection of dust up to about 20 kpc perpendicular to the disk of edge-on galaxies. With a statistical approach, \citet[][MSFR]{menard10} measured the cross-correlation between the colors of distant quasars and foreground galaxies as a function of the impact parameter to galaxies. They found an excess reddening signal on scales ranging from 20 kpc to a few Mpc, implying that the distribution of dust extends all the way to the intergalactic medium. \citet{2012ApJ...754..116M} showed that a similar amount of dust can been seen associated with strong Mg II absorbers at $0.5 < z < 2.0$, providing another line of evidence of the presence of dust on large scales around galaxies. \citet{Fukugita2011} showed that the summed contributions of dust in and outside galaxies appears to be in agreement with the total amount of dust that ought to be produced in the Universe. This implies that dust destruction does not play a major role in the global dust distribution and that most of the intergalactic dust survives over cosmic time. 

In this work we focus on the low-redshift Universe and constrain both the amount of dust surrounding nearby ($z \sim 0.05$) galaxies and its dependence on galaxy properties. We do so by using a set of standard crayons, following the work of \citet{2010ApJ...719..415P}, i.e. galaxies for which the colors can be standardized. The outline of the paper is as follows: in \S \ref{obs} we describe the datasets how we estimate the color of a standard crayon galaxy. In \S \ref{methods} we introduce our estimator for reddening measurement, in \S \ref{aar} we present our analysis and results showing both detections of reddening and the variation of this reddening as a function of foreground galaxy parameters. We discuss these results in the context of galaxy formation and metal budgets in \S \ref{discussion}, and conclude in \S \ref{conclusion}.

\section{Data}\label{obs}

Our goal is to measure the mean color change of distant sources induced by the presence of foreground galaxies, as a function of impact parameter, $R$. Optimizing such a measurement requires both maximizing the number of foreground-background pairs and minimizing the possible scatter in the color distribution of the background objects. To do so we work with data drawn from the Sloan Digital Sky Survey \citep[SDSS;][]{york00}.

For our foreground population we select galaxies from the Main Galaxy Sample from the data release 7 \citep{strauss02} and use the magnitude limited selection $r < 17.77$. This produces a sample of 695,652 foreground galaxies for which two intrinsic properties are extracted from the MPA-JHU value-added catalog \citep{Kauffmann:2003iz, Brinchmann:2004hy}: specific star-formation rate (sSFR) and stellar mass ($M_\star$). The distribution of galaxies as a function of redshift is shown in the left panel of Figure \ref{galz_cdisp} with the black histogram. 

For our background sample we make use of standard crayons, as introduced by \cite{2010ApJ...719..415P}: objects whose colors have a very narrow distribution. We select passively evolving galaxies from the Main Galaxy Sample (again, $r < 17.77$) using the criterion that they have neither detectable O{\sc ii} nor H$\alpha$ emission, significantly limiting their number. The details of that selection procedure are discussed in PG10. This provides us with a sample of 151,637 galaxies. We note that we do not exclude these galaxies from our foreground sample described above. The redshift distribution of the background galaxies is shown in the left panel of Figure \ref{galz_cdisp} in gray.

To estimate color changes we use the apparent model magnitudes of the galaxies and apply two corrections to them. First, we correct for the effects of Galactic dust reddening. To estimate them we avoid using the standard dust map from \cite{SFD98} which relies on FIR emission which is known to originate not only from the Milky Way but also from low redshift galaxies (see \citei{Yahata07}, \citei{Kashiwagi:2012we}, \citei{Peek:2013eh}, for a discussion of the effect). Instead, we emulate the work of \citet{Burstein:1982dz} and use a Galactic dust map based on the distribution of HI, as probed by the the Leiden-Argentina-Bonn radio survey \citep{Kalberla05}. We define the Galactic dust reddening $\Delta_{\rm MW}$, measured in the bands $a$ and $b$, in a given direction in the sky by
\beq\label{HIdust}
\Delta {\rm C_{\rm MW}} = \left(R_a-R_b\right) \frac{N_{\rm HI}}{7 \times 10^{21} \rm cm^{-2}},
\eeq
where $R_a$ and $R_b$ are the Galactic extinction coefficients for the SDSS bandpass filters \citep{Stoughton:2002dg}, and the ratio of reddening to \hi for this region is derived from \cite{2013ApJ...766L...6P}. We exclude galaxies with a measured $E\left(B-V\right) > 0.1$, where \hi column is known to be a poor reddening estimate, removing 1891 galaxies from our background sample. We note that using the \cite{SFD98} extinction rather than the method described above does not qualitatively modify our conclusions. Second, we need to remove possible trends between background galaxy colors and redshifts, which stem both from galaxy evolution and ``K-correction''. To do so we measure the median color-redshift relation for each of the 10 SDSS optical colors and characterize it using a fourth-order polynomial fit, refered to as $\Delta {\rm C_{\rm redshift}}$.

With these two correction terms in hand we finally define the corrected color of a galaxy by
\beq\label{cprimeab}
{\rm C_{\rm corr} = C_{obs} - \Delta C_{\rm MW} - \Delta C_{\rm redshift}}\;.
\eeq 
This provides us with a distribution of corrected colors ${\rm C_{\rm corr}}$ for which the dispersion ranges from 21 to 180 millimagnitudes depending on the chosen bands. The right panel of Figure \ref{galz_cdisp} shows examples of such distributions. 

\begin{figure*}[!ht]
\includegraphics[scale=0.8]{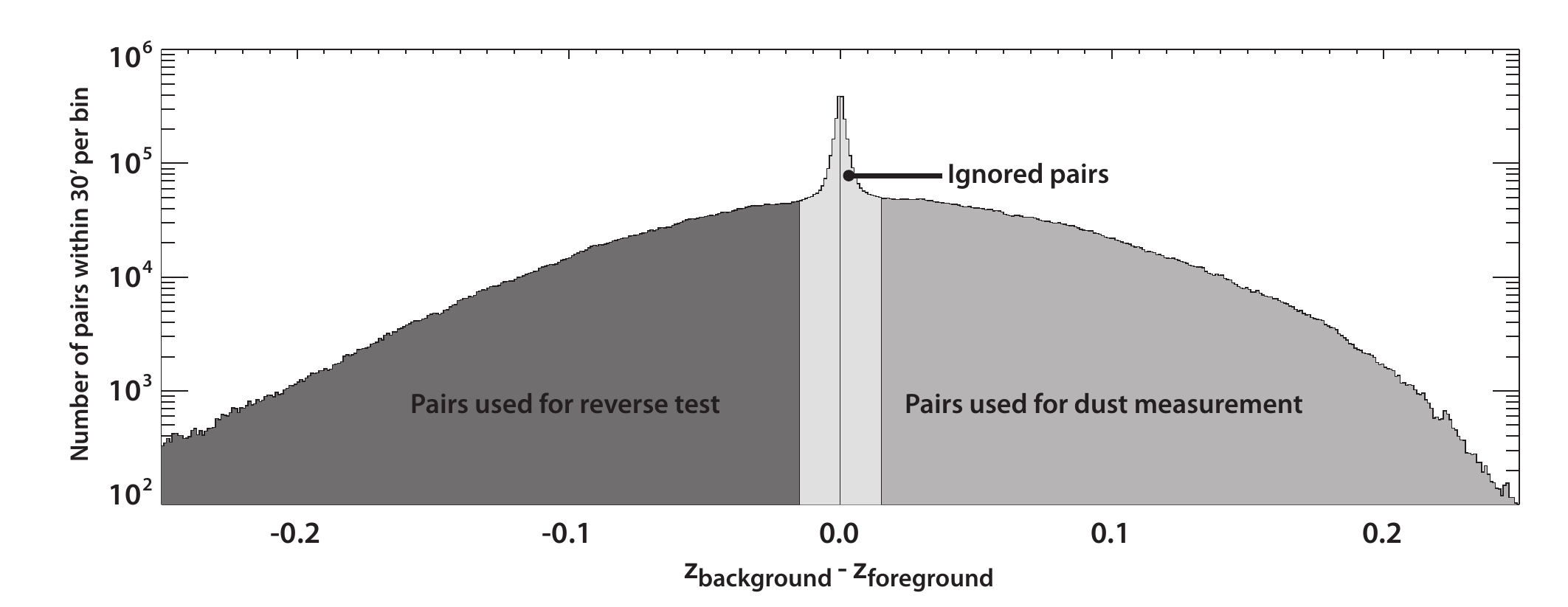}
\caption{Distribution of redshift differences for the galaxy pairs selected within 30 arcminutes. The light gray region shows pairs used for the direct dust reddening measurements. The dark gray region shows pairs used for the ``reverse test" used to characterize the effects of photometric bias, where the "background" galaxy is \emph{in front of} the ``foreground'' galaxy. The lightest gray area at the center represents pairs that are likely to be related in physical space ($| z_{\rm background} - z_{\rm foreground}| < 0.015$) and are thus not used in either measurement.}
\label{hzb}
\end{figure*}

\subsection{Outlier Rejection}

To increase the robustness of our statistical analysis we first remove outliers in the color distribution of the background galaxies (PG10, figures 2 \& 3). To do so we use Chauvenet's criterion, rejecting outliers that have larger deviations in color than would be expected for any points, given the standard deviation of the sample and the number of galaxies in the sample, in our case 4.35$\sigma$.  We reject galaxies that fail to meet this criterion in any of the 10 colors, which removes 7,861 galaxies from the sample, i.e. about 5\% of the data. We note that arbitrarily moving the threshold such that we reject half or twice the number of galaxies, or simply clipping at 3 $\sigma$ does not qualitatively effect our results. We also test against a bootstrap analysis and do not detect differences in the measured errors. Our extreme outlier rejection criterion is indicated as vertical dashed lines in the right panel of Figure \ref{galz_cdisp}. The final sample contains 141,885 standard crayon galaxies.


\section{Methods}\label{methods}

Given that the expected level of CGM dust reddening is lower than the intrinsic width of the corrected color distribution, we constrain its amplitude by measuring the mean color excess induced by the presence of foreground galaxies in a range of impact parameters, $\langle {\rm C}(R_{min},R_{max})\rangle$, where $R_{min}$ and $R_{max}$ are the minimum and maximum impact parameter. In addition to reddening by dust, apparent color changes can stem from other effects and systematic errors: physical correlations between foreground and background objects (galaxy clustering), distortions in the photometry due to the presence of a nearby galaxy and wide-field photometric errors. Below we define an estimator which takes those effects into account.

One requirement we put on our estimator is that it should not lead to any reddening signal if we measure the color changes of \emph{foreground} galaxies as a function of angular separation from background galaxies. Any such change would indicate our estimator is sensitive to some non-physical effect, as light coming from a galaxy in front should not be modified physically by a galaxy behind. We refer to this measurement as the ``reverse test".

\subsection{Physical and lensing-induced correlations}

Galaxies are known to follow a color-density relation \citep[e.g.][]{Blanton:2005eb} leading to mean color change as a function angular separation. To avoid being contaminated by such an effect, we impose a minimum redshift difference between our background and foreground objects. In Figure \ref{hzb} we show the distribution of foreground-background pairs within 30 arcminutes as a function of $\delta z = z_{\rm background}-z_{\rm foreground}$ for both the direct reddening measurement (right) and the reverse test (left; see \S \ref{sspe}). There is a sharp increase in the number of pairs on small scales, due to the clustering of galaxies. To be conservative we ignore galaxy pairs with $|\delta z| < 0.015$. Changing this value by a factor of 2 does not qualitatively effect our results. 

Gravitational lensing increases the brightness of background galaxies behind foreground galaxies. Because of this brightening, there will tend to be more background galaxies brighter than a given threshold in the vicinity of foreground galaxies \citep{Narayan:1989bb}. The strength of this effect depends on both the gravitational magnification of the foreground population and the shape of the brightness distribution of the background sources. While our results are insensitive to background galaxy brightness, the average color of a population of galaxies will slightly change if the limiting magnitude is effectively altered. This effect, similar to the population reddening effect discussed in \cite{Peek:2013eh}, is very small; for our sample less than 20 millimagnitudes of color change per magnitude of brightening for all the colors we investigate. On scales of 150 kpc (about three arc\-minutes at the redshifts of interest), typical changes in galaxy brightness are $\sim 3 \times 10^{-3}$ and drop with scale roughly as $R^{-1}$ (Scranton 2005, MSFR).  We therefore expect lensing-induced color changes to be $\sim 5 \times 10^{-5}$ mag at R $\sim$ 150 kpc. This term is insignificant and we neglect it in the following.

\subsection{Wide-field photometric errors}

We expect wide-field variations in the photometry and colors at some level, as the SDSS survey was conducted over varying conditions over many years. This can be induced by calibration offsets \citep{Padmanabhan:2008js} or biases in the dust map \citep[e.g.,][]{2013ApJ...766L...6P}. Such large-scale variations may bias the color change we wish to measure: large scale structure in our foreground galaxies may overlap with regions of photometric error, generating a bias beyond simple Poisson noise. In order to mitigate this effect, we only focus on the excess color change measured with respect to a large-scale averaged mean color:
\begin{eqnarray}
\Delta {\rm C}(R_{min}, R_{max}) &=& 
\langle {\rm C_{\rm corr}}(R_{min}, R_{max}) \rangle \\
&& \,-\, \langle {\rm C_{\rm corr}}(1\,\rm Mpc,2\,\rm Mpc) \rangle \nonumber
\end{eqnarray}
We find this effect to be typically of order $ 2 \times 10^{-4}$ magnitudes, and always smaller than our error bars and signal. We note that this method restricts us to measuring dust with $R < 1 $ Mpc and represents an imperfect mitigation of this very small effect.

\subsection{Small-scale photometric biases}\label{sspe}

On small scales, photometric estimation can be affected by the presence of nearby extended sources, like bright galaxies \citep{Aihara:2011kj}. The foreground galaxy is not a point source, and thus may have a real extended light distribution, or scattered light. The fact that the background object is also a galaxy introduces additional opportunities for error. To consistently measure the apparent magnitude of a galaxy, one must fit a model to the galaxy profile, which requires a careful estimate the sky brightness nearby. Any other galaxies in the same region of sky can bias the sky brightness estimate.

Quantifying such effects as a function of galaxy proximity and brightness requires tests of the SDSS photometric pipeline (see \citei{Huff:2013bc} and future work discussed therein). Instead of attempting to predict the amplitude of these effects, we can directly estimate them using the reverse test introduced above, i.e. measuring the color changes of \emph{foreground} galaxies as a function of angular separation from \emph{background} galaxies. Such a correlation should not lead to any signal. Therefore any measured quantity is an estimate of the amplitude of photometric biases. We perform a photometrically matched measurement $\Delta C_{\rm reverse}$.
To do so we weight galaxy pairs in the reverse test such that their pair-wise angular distribution matches that of the forward measurement. Our final circumgalactic dust reddening estimator is thus
\beq
\label{debias}
{\rm \Delta C_{\rm dust} = \Delta C - \Delta C_{\rm reverse}}\;.
\eeq
This small-scale photometric bias is expected to depend on the brightness of the foreground galaxies. Since the reverse test will tend to have higher redshift ``foregrounds'' than the dust measurement, we need take this difference into account. We estimate $\Delta C_{\rm reverse}$ as a function of the magnitude of ``foreground'' galaxies by binning the sample by quartiles in $r$ band magnitude. We find that, in practice, the strength of this small scale photometric bias is insensitive to whether we bin by quartiles or in some other way, and whether we use $r$ or another filter. A comparison of $\rm \Delta C_{\rm dust}$ and $\rm \Delta C$ is shown in the following section.

\section{Results}\label{aar}

\subsection{Mean reddening signal}\label{mrs}

\begin{figure}[!t]
\begin{center}
\includegraphics[scale=1.1]{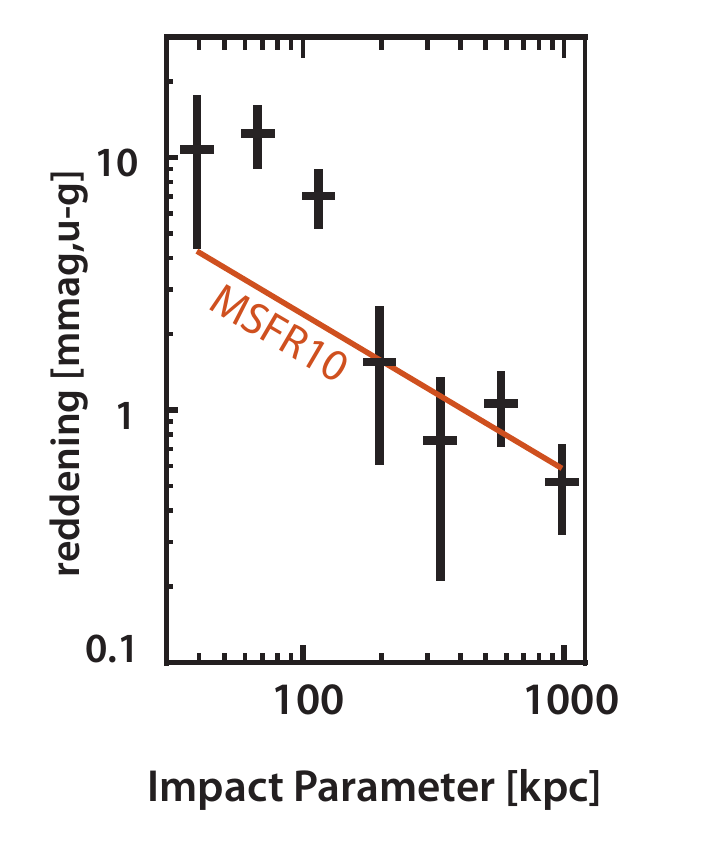}
\caption{Mean $u-g$ reddening excess measured as a function of scale, indicating the presence of dust in the CGM and IGM. The red line shows the best fit power law obtained by MSFR who measured the reddening excess induced by $z = 0.3$ galaxies on background quasars. Our analysis indicates, for the $u-g$ color, a higher level of reddening within 150 kpc and a departure from a power law. As shown in Fig.~\ref{color30--150} this excess is not seen in the other bands.}
\label{radialtrends}
\end{center}
\end{figure}

Using the color excess estimator presented in Equation \ref{debias} we measure the mean reddening induced by foreground galaxies for as a function of impact parameter, from 30~kpc to 1~Mpc. In Figure \ref{radialtrends} we show the radial profile of the $u-g$ color excess for 7 radial bins. We detect a reddening signal of about 10 milli-magnitudes at $R_p < 150$~kpc, which drops to about 1 millimagnitude at a scale of 1 Mpc. We attribute this signal to the presence of dust in the CGM and IGM. For comparison we show as a red line the best fit power-law trend to the results obtained by MSFR using foreground galaxies at $z\sim0.3$ and background quasars to measure colors. Overall, our results show a similar effect. This illustrates the power of standard crayons to perform low-level reddening measurements by reducing the intrinsic scatter in colors rather than increasing the size of the background sample. Our measurement also shows some differences with respect to the MSFR best fit trend. We observe an enhanced signal on scales ranging from 30 to 150 kpc.

\begin{figure}[h]
\begin{center}
\includegraphics[scale=.9]{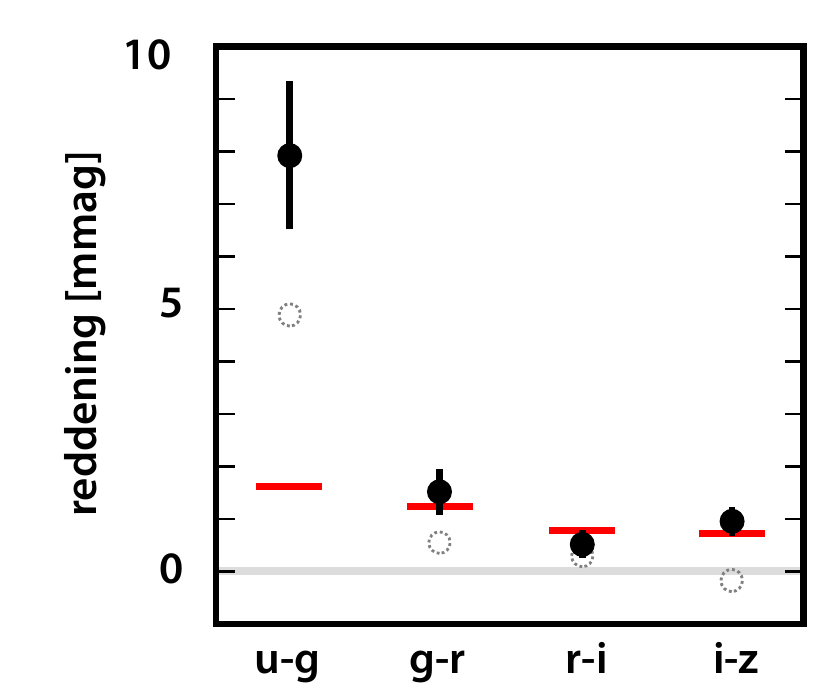}
\caption{The average reddening of background galaxies with impact parameters between 30 and 150 kpc of all spectroscopic foreground galaxies. Solid black circles represent our dust reddening estimator $\Delta C_{\rm dust}\left(30~ \rm kpc, 150~ \rm kpc\right)$. An excess reddening is detected in all color combinations. The red lines represent the results of MSFR, assuming SMC dust. For reference, the empty gray circles show the results without the reverse test debiasing, $\Delta C\left(30~ \rm kpc, 150~ \rm kpc\right)$.}
\label{color30--150}
\end{center}
\end{figure}

In the regime $30$ kpc $<R_p<150$ kpc, where the signal-to-noise ratio of our measurements is higher, we can investigate the wavelength dependence of the measured color change. This is shown in Figure \ref{color30--150}. Our measurements show that the wavelength dependence of the signal is generally consistent with that of an SMC bar exctinction curve \citep[red dashes,][]{2001ApJ...548..296W} except at the shortest wavelengths. The $u-g$ color shows a significant excess, inconsistent with the SMC bar reddening curve, with a reduced $\chi^2$ = 7.6. We note that our $u$-band extinction is not affected by the canonical 2175\AA (NUV) absorption ``bump'' at the redshift of our foreground galaxies, i.e. $0.03<z<0.1$.

In general, small dust grains contribute to a steeper slope on the blue end of an extinction curve, generating an increase in $u-g$ reddening. To test the hypothesis that the increase in $u-g$ color relative to SMC extinction (Figure~\ref{color30--150}) is driven by the dust grain size distribution, we examine the very simplified case in which the dust has a single grain size and material. We do not expect that CGM grains are truly uniform, but we use this toy model as a simple test appropriate for our broadband data.

We investigate two extremely simplified dust models: graphite only and silicate only, each with a single variable grain size $a$.  
We use dielectric constants given by \cite{Draine2003b}, \cite{DL1984}, and \cite{LD1993} and compute Mie scattering and extinction cross-sections using the publicly available \texttt{bhmie} code \citep{BH1983}.

While both the graphite and the silicate models show a best fit for smaller grains, $a_g = 0.05~\mu{\rm m}, a_s = 0.06~\mu{\rm m}$, both fall short of the u-band extinction, with reduced $\chi^2$ of 6.5 and 5.5, respectively. Therefore, we cannot conclude that we have strong evidence for a small grain population in the CGM. More likely, there exists some weak, unmodeled systematic that is increasing our errors. We also fit the data with the SMC curve but excluding the $u$-band and we find a better fit with reduced $\chi^2$=2.3. For the remainder of this work we exclusively refer to the simpler SMC dust assumption. 

We can derive an overall CGM dust mass using
\beq
\label{dust-mass}
M_{\rm dust} = \int_{30~ \rm kpc}^{150~ \rm kpc} \frac{1.086\ A}{\kappa} 2 \pi r dr
\eeq
where $A$ is the extinction expected given the measured SMC reddening and $\kappa$ the SMC dust opacity. The fit both using and excluding the $u$-band data show a dust mass of $6 \pm 2 \times 10^7 M_\odot$; the fit is largely insensitive to $u$ as the $u$-band errors are large. These results are consistent with the result of MSFR, who found a CGM SMC dust mass of $5\times 10^7 M_\odot$. 

\subsection{Trends with galaxy properties}\label{galprop}

\begin{figure}[h]
\includegraphics[scale=0.9]{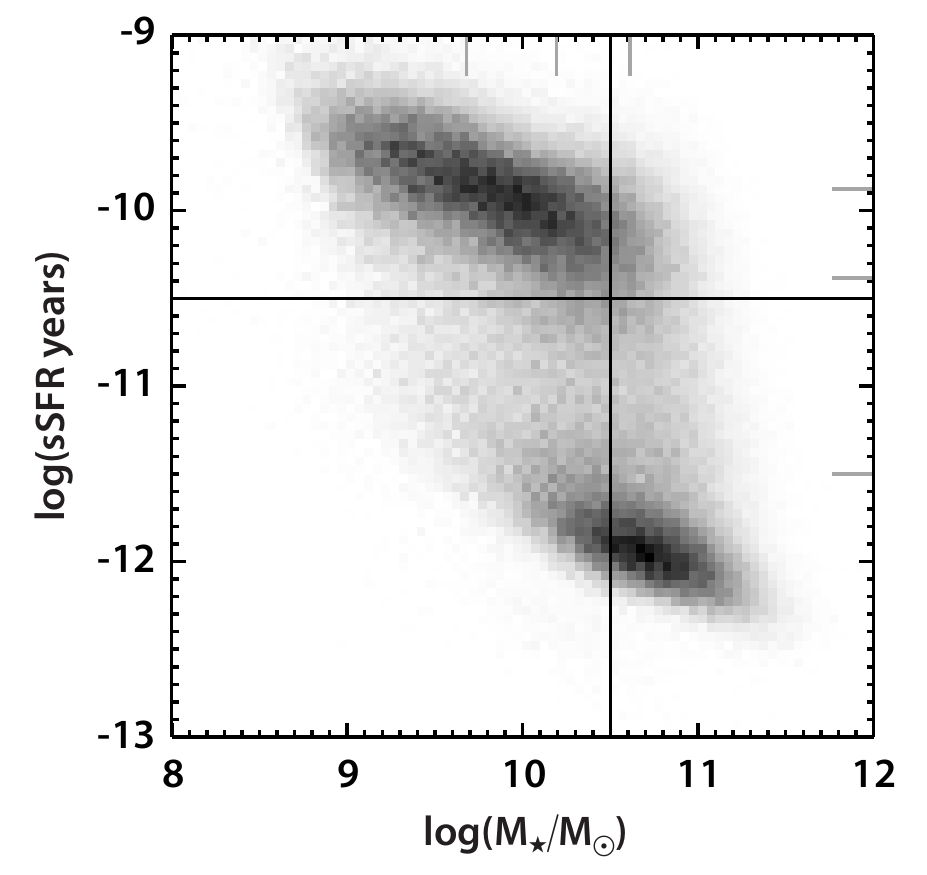}
\caption{The bivariate distribution of foreground galaxies in sSFR and $M_\star$. The univariate quartile values are shown with long gray tickmarks; the splits used in \S \ref{galprop} are shown in black.}
\label{histofg}
\end{figure}

\begin{figure}[h]
\includegraphics[scale=1.0]{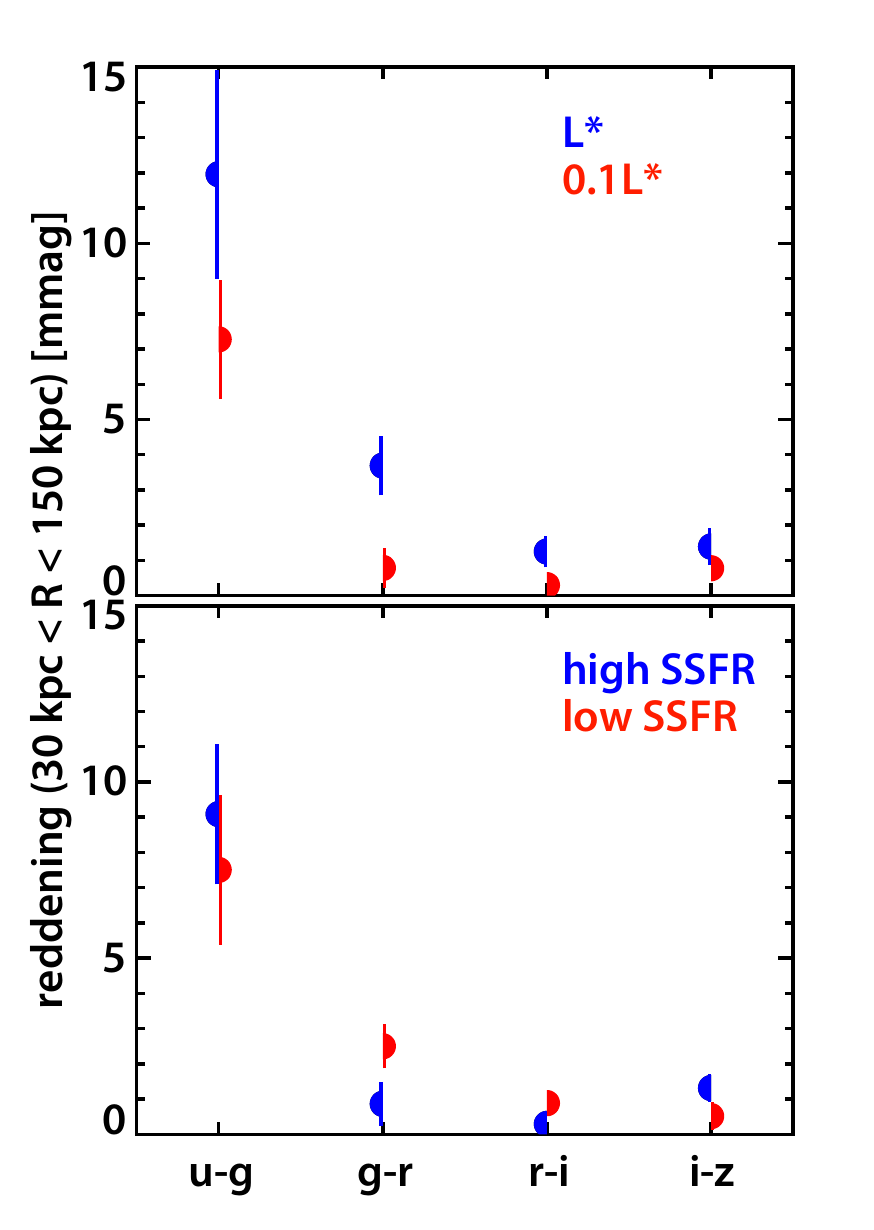}
\caption{Average reddening between 30 and 150 kpc impact parameter of foreground galaxies, $\Delta C_{\rm dust}\left(30~ \rm kpc, 150~ \rm kpc\right)$, split into sub-samples by stellar mass (top panel) and specifc star formation (bottom panel). In the top panel the blue half-circles and errors represent the measurement of a sub-sample with average stellar mass of L* (the L* sample), while the red half-circles and errors represent the sub-sample with average stellar mass of 0.1 L* (the 0.1 L* sample). In the bottom panel the blue half-circles represent the high specific star formation rate sub-sample, at an average of 2 $\times 10^{-10}$ yr$^{-1}$, while the red half-circles and errors represent the low specific star formation rate sub-sample, at an average of 8 $\times 10^{-12}$ yr$^{-1}$. It is visually evident that the higher mass galaxies produce more reddening, but not ten times as much, as would be expected if CGM dust mass scaled linearly with galaxy stellar mass. No reddening trend is detectable as a function of sSFR.}
\label{overallmass_ssfr}
\vspace{0.5cm}
\end{figure}

\begin{figure}[!t]
\includegraphics[scale=0.8]{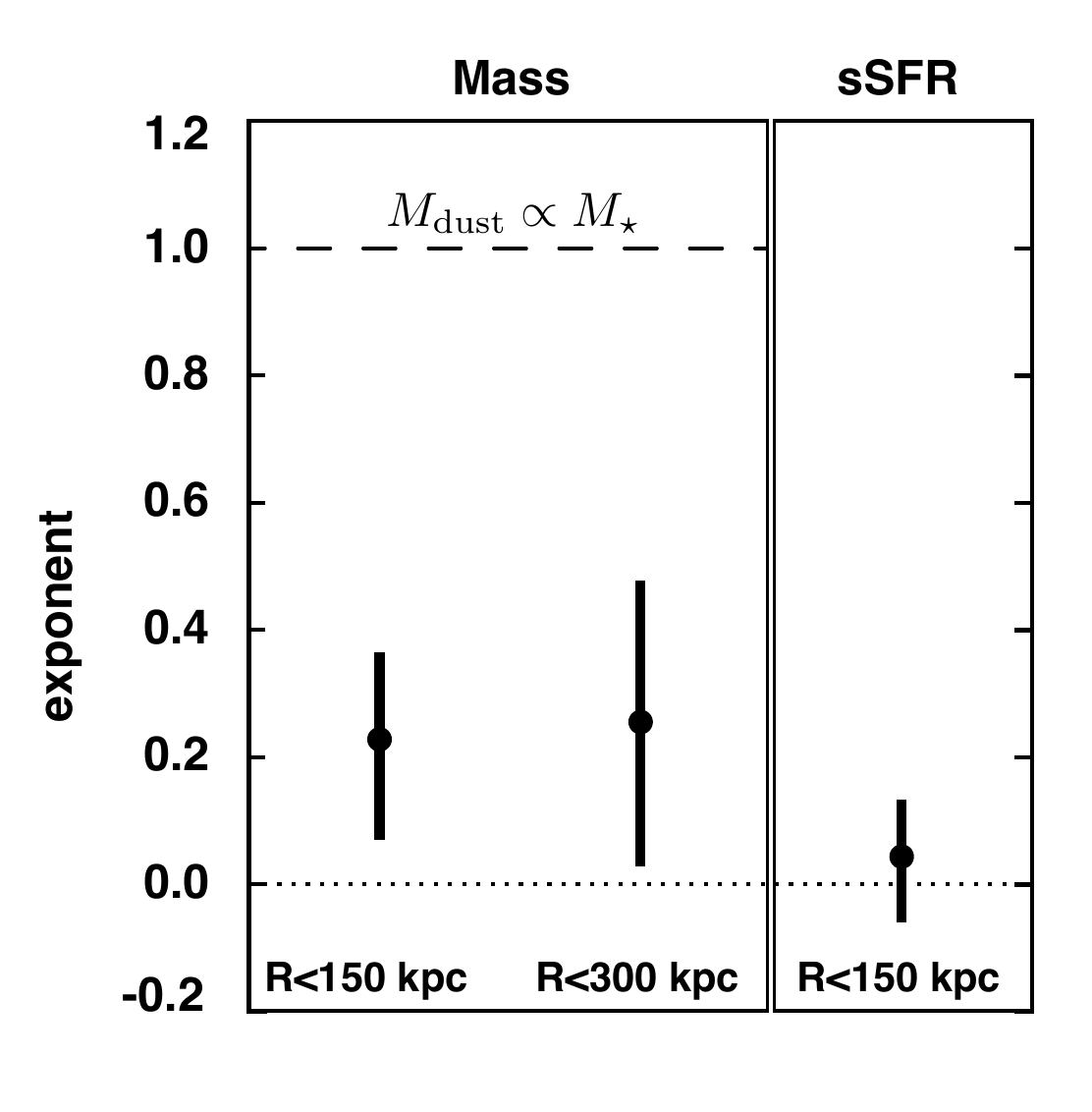}
\caption{The dependence of circumgalactic reddening on galaxy properties. The left panel shows the exponent $\beta$ characterizing the dependence on stellar mass $M_{\rm dust} \propto {M_\star}^\beta$, as a function of scale. The right panel
shows the exponent $\gamma$ characterizing the dependence on specific star formation rate $M_{\rm dust} \propto {sSFR}^\gamma$.}
\label{mass}
\end{figure}

Having access to a number of physical parameters for the foreground galaxies we can investigate correlations between the amount of dust in the CGM and galaxy properties. To do so we split the foreground sample by mass at $M_\star = 3 \times 10^{10} M_\odot$ (about 1/2 L*; see Figure \ref{histofg}). The average stellar mass of the low-mass group is $6 \times 10^9 M_\odot \simeq 0.1 $L* while the average for the high-mass group is $6 \times 10^{10} M_\odot \simeq $L*. We call these groups the $0.1$L* and L* sub-samples, respectively. 

We determine the CGM reddening for each subsample and their ratios, shown in Figures~\ref{overallmass_ssfr} and \ref{mass}, respectively. In determining the ratio we make the explicit assumption that ratio of reddening to dust column is not a function of physical parameter by which we split the sample. We detect more reddening around L* galaxies over our fiducial impact parameter range (30-150 kpc), but only by a factor of $1.7_{-0.5}^{+0.7}$. If we assume a power-law dependence of CGM dust mass on host stellar mass we find
\begin{align}
\label{beta}
    &M_{\rm dust} \propto {M_\star}^\beta 
    \rm{~~with~~}\beta = 0.23_{-0.15}^{+0.15}.
\end{align}
The ratio does not change significantly when we increase the range of impact parameter to 300 kpc, indicating that a more extended dust distribution around L* galaxies does not explain this weak dependence of CGM reddening with galaxy mass.
We rule out a linear proportionality between $M_{\rm dust}$ and $M_\star$ at a level greater than 4$\sigma$.

We repeat the experiment, splitting the galaxy population by sSFR at 3 $\times 10^{-11}$ yr$^{-1}$ (Figure~\ref{histofg}). The low sSFR subpopulation has an average sSFR of 8 $\times 10^{-12}$ yr$^{-1}$ while the high sSFR subpopulation has an average sSFR of 2 $\times 10^{-10}$ yr$^{-1}$. The trend with sSFR is characterized by
\begin{align}
\label{gamma}
    &M_{\rm dust} \propto {\rm sSFR}^\gamma 
    \rm{~~with~~}\gamma = 0.04_{-0.09}^{+0.10}\;.
\end{align}
This is also reported in Figure \ref{mass}. It indicates that star forming and passively evolving galaxies around surrounded by a similar distribution of dust in their halo.

\subsection{Trends with disk angle}

\begin{figure}[!t]
\includegraphics[scale=1.2]{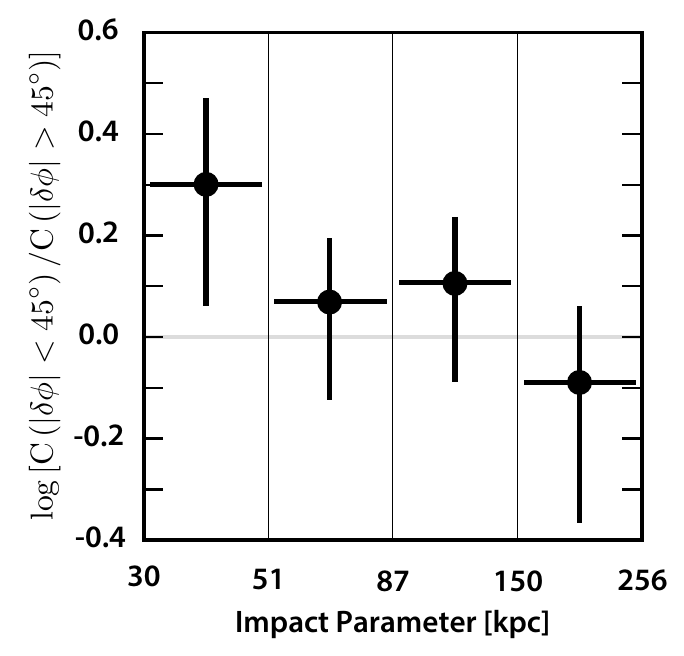}
\caption{The dependence of circumgalactic reddening on position angle around the galaxy as a function of impact parameter. The x-axis shows five radial bins from the forground galaxy, the y-axis shows the ratio of reddening seen in galaxies with a background along the minor axis to the reddening seen in galaxies with a background along the major axis. We only a faint signal within 50 kpc, excluding the null at 91\% confidence.}
\label{phirat}
\vspace{0.5cm}
\end{figure}

Finally, we explore the spatial distribution of CGM dust as a function of position angle with respect to the major axis of the galaxy. To measure this, we first find a subsample of the target galaxies where the orientation of the major axis (position angle) of the galaxy is well measured. The SDSS photometric pipeline measures three position angles: the $\phi_{exp}$ of the exponential fit, the $\phi_{deV}$ of the deVaucouleurs fit, and the $\phi_{iso}$ of the best-fit ellipse to the 25 magnitudes per square arcsecond isophote. In cases where the galaxy's major axis is difficult to discern, these three fits will report essentially arbitrary values for $\phi$. To eliminate such cases, we consider only galaxies that have $|\phi_{iso} - \phi_{exp}| < 15^\circ$, $|\phi_{iso} - \phi_{deV}| < 15^\circ$, and $|\phi_{deV} - \phi_{exp}| < 15^\circ$, which is about 68\% of the total sample. We then measure the position of each background galaxy with respect to the $\phi_{dev}$ of the foreground galaxies to determine an azimutal angle $\delta\phi$ for each pair. Finally we bin the pairs into two groups: background sources probing the major axis region of the foreground galaxies ($|\delta\phi| < 45^\circ$) and those probing the minor axis region ($|\delta\phi| > 45^\circ$). We measure the ratio of reddening along the minor axis ($|\delta\phi| < 45^\circ$) to reddening along the major axis ($|\delta\phi| > 45^\circ$), as a function of impact parameter. We find an indication of an excess of reddening signal along the minor axis, excluding the null with 91\% confidence, compared to the major axis, on scales ranging from 30 to 50~kpc, shown in Figure \ref{phirat}. No significant difference is seen on larger scales.

\section{Discussion}\label{discussion}


Characterizing dust reddening effects at a level below one percent is challenging. A number of effects unrelated to dust can bias the measurements: physical clustering, large-scale photometric variations, offsets in the estimation of the sky background level near extended sources, gravitational lensing-induced color changes due to possible brightness-color trends. In this analysis we have accounted for these effects by using an estimator sensitive the excess reddening as a function of angular separation and only accounting for color changes affecting background sources. Our results provide a new line of evidence that a substantial amount of dust resides in galactic halos. As opposed to previous statistical studies using quasars as background sources \citep[e.g.][]{Chelouche:2007dy,menard10} we have shown that it is possible to use background galaxies, and in particular standard crayons \citep{2008ApJ...688..198B,2010ApJ...719..415P}.

\subsection{The Spatial Distribution of CGM dust}

The increased reddening detected within 150 kpc (Figure \ref{radialtrends}) appears at first to be inconsistent with the smooth power-law found by MSFR.  However, the MSFR result assumes that all of the foreground galaxies are at $z = 0.34$.  As the MSFR results are based on photometrically detected foreground galaxies over a wide range of redshifts, we expect that any radial CGM edge in physical space would be washed out by the uncertainty in redshift.

The increased detection within 150 kpc is qualitatively consistent with absorption lines of highly ionized metals \citep{2009ApJS..182..378W} and cooler metals in the CGM \citep{Bordoloi:2011iu}. We find further similarity with gas-phase metals in the position angle distribution. The hint of increased reddening along the minor axis at low impact parameters is also consistent with previous work constraining the distribution of MgII absorption around galaxies \citep{Bordoloi:2011iu,2014arXiv1404.5301L}. These measurements of dust spatial distribution present an interesting and coherent picture to test against galaxy formation scenarios.

\subsection{Trends with Galaxy Properties}

A new frontier reached in this analysis is the ability to probe the relation between the amount of CGM dust and galaxy properties. Our results indicate that $M_{\rm dust, CGM} \propto M_\star^{\beta}$ with $\beta\sim0.2$ (Equation \ref{beta}). This trend provides us with a new constraint for the modelling of galactic winds \citep{Zu:2010ch,Oppenheimer:2006eq,Murray:2004jt} and metal pollution on large scales. We note that recent studies of CGM metals through absorption lines analyses also point out to a weak dependence between with galactic stellar mass \citep{Werk:2013ir,Zhu:2013bs}.

We note that the relatively weak relationship 
between the amount of CGM dust and galaxy stellar mass
found in Equation \ref{beta} cannot persist to lower mass dwarf galaxies. 
\citep[][;P14]{Peeples:2014fe} includes an analysis of the total mass of metals produced by galaxies as a function of mass. Extrapolating their relation -- that galaxies produce $\sim5\%$ of their stellar mass in metals (figure 1) -- we find that our Equation~\ref{beta} predicts more mass in CGM dust than is produced by a galaxy of $M_\star \simeq 5 \times 10^8 M_\odot$. This is of course unphysical; there must be some steepening of this relationship toward lower mass. 

Lastly, the lack of detectable relationship between sSFR and CGM dust content has implications for dust lifetime in the CGM and feedback mechanisms. Our blue sample has a sSFR 25 times higher than the red sample, and yet we see no effect on the reddening from CGM dust. If dust is primarily ejected from galaxies by processes that are connected to star formation, it must survive for gigayears in the CGM to be equally represented around starforming and passively evolving galaxies.

\subsection{The Metal Budget}

\begin{figure}[h]
\includegraphics[scale=1.0]{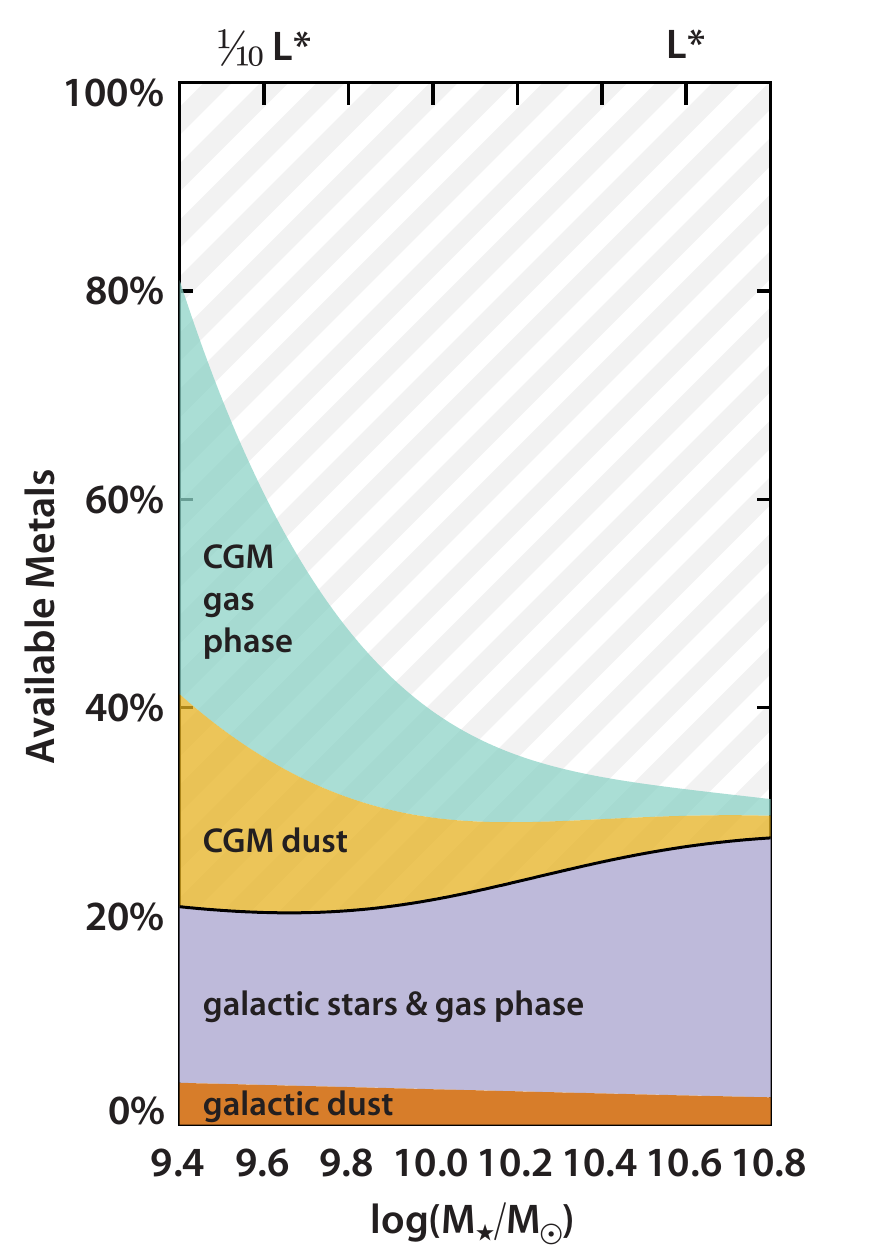}
\caption{An accounting of all available metals as a function of galaxy mass, from the analysis of P14. Below the black line are metals in stars and the gas-phase ISM (purple), and ISM dust (orange). Above the black line are CGM metals: low-ions and OVI-traced metals (green) and CGM dust from this work (yellow). The hashed area represents metals missing from galaxies themselves. There is a very clear difference in the fraction of metals accounted for in 0.1 L* galaxies versus L* galaxies. For a discussion of uncertainties on these values see P14.}
\label{molly}
\end{figure}

Determining the location and state of metals can help us understand how galaxies produce and expel material, and therefore how they evolve. While early work highlighted the location of metals at high redshift \citep[e.~g.~][]{2007MNRAS.378..525B}, recently a complete analysis has been done by P14. This work determined the total mass of metals formed by a galaxy by $z \sim 0$ as a function of $M_\star$ that were not immediately locked in stellar remnants \citep{Fukugita:2004bf}: the ``available metals'' and discussed whether how much were detected within galaxy disks or in the CGM.

To illustrate the significance of CGM dust to this budget, we show a new version of P14 figure 9 including our measurement of the very weak dependence of CGM dust mass on host galaxy stellar mass (Figure \ref{molly}), under the assumption from \cite{menard10} that a galaxy of stellar mass $10^{10.4} M_\odot$ has CGM dust mass of $5 \times 10^7 M_\odot$. A number of assumptions go into this Figure which are discussed at length in P14. Importantly, P14 assumes no dependence of CGM gas-phase metal mass on $M_\star$ ($\beta = 0$). However, observations of OVI and lower ions can only rule out a linear, or stronger, dependence on $M_\star$ (rule out $\beta \ge 1$). Part of this uncertainty stems from the relatively few low-mass galaxies observed in the COS-Halos data set: only four in the range $4 \times 10^{9} M_\odot \lesssim M_\star \lesssim 10^{10}  M_\odot$ \citep{Tumlinson:2013cl}. Our dust work presents a more precise measure of $\beta$ (Equation \ref{beta}), in part due to the much larger range of foreground galaxy masses probed. 

While each of these $\beta$ measurements are subject to various independent uncertainties and biases, the fact that all three measurements (OVI ions, lower ions, and dust) show weak dependence on $M_\star$ in this mass range suggests it is worth examining the physical implications. With these fiducial assumptions, a large fraction of the metals produced by 0.1 L* galaxies reside in the CGM, weakening any ``missing metals'' problem at that mass. Indeed, a 0.1 L* galaxy contains approximately as much metal in CGM dust as in the entire galaxy itself. Conversely, very little of the available metals from L* galaxies are found in detected phases of the CGM, suggesting that the missing metals must be in some unprobed phase or outside the CGM. These results give a valuable perspective on galaxy formation models, especially in the context of galactic feedback.

\section{Conclusion}\label{conclusion}

We can draw several conclusions from this work.
\begin{itemize}
\item Using galaxies and in particular standard crayons as background sources we detect the excess reddening induced by the extended distribution of dust around galaxies.
\item{We confirm the existence of $\sim 5 \times 10^7 M_\odot$ of dust in the CGM of 0.1 L* -- L* galaxies, consistent with the results of MSFR based on reddening of background quasars.}
\item We find a weak dependence of CGM dust reddening on galaxy stellar mass
$M_{\rm dust} \propto {M_\star}^{0.2}$ (Equation \ref{beta}), and no detectable dependence on specific star formation rate.
\item Including constraints on the distribution of metals from absorption line studies, the dust contribution to the overall metal budget indicates that the missing metals problem at low redshift is more acute near L* galaxies than near 0.1 L* galaxies.
\end{itemize}
We have found in this work that the photometry of close pairs of galaxies is susceptible to significant systematic error, even in a survey with photometry tested as meticulously as the SDSS \citep{Padmanabhan:2008js}. To press forward with the measurement of extragalactic dust reddening, we must build photometric pipelines that are resistant to the kinds of biases discussed in this paper. Future ground-based surveys and space missions may sidestep some issues with more stable seeing, darker skies, and higher resolution.  However, without a clear specification to deliver photometry unbiased by close neighbors, we cannot assume that such surveys will be optimal. Better bias constraints on the photometry of close pairs are also very important for weak lensing magnification measurements \citep{Huff:2013bc}. An alternative is to use observations of color-standardized point sources (quasars) to study extragalactic reddening.

With these important caveats in mind, we look toward the future of measuring dust in the CGM of galaxies with spectroscopic galaxy data. Expanding the analysis to a broader range of wavelengths would help better constrain the dust extinction properties, perhaps including WISE \citep{2010AJ....140.1868W} and GALEX \citep{2005ApJ...619L...1M} data. A preliminary investigation has shown much stronger systematic biases in close galaxy pairs when comparing photometry across surveys. Data from the GAMA survey \citep{Driver:2011gp} may help us reach higher precision, especially because the multi-pass spectroscopic survey avoids fiber collision issues, and many more close pairs can be studied. BOSS \citep{Dawson:2012fj} has observed ten times more passively evolving galaxies than the original SDSS data, but its targets are selected with cuts in color space, which may contaminate the standard crayons. If this contamination can be well understood, BOSS may allow us to study CGM reddening at $z \sim 0.5$. Future surveys like the notional high latitude survey proposed for WFIRST \citep{2013arXiv1305.5425S} may allow similar measurements toward $z \sim 2$, at the height of galaxy formation in the history of the universe.\\

\acknowledgements

We thank Daniel Rabinowitz of the Columbia statistics department for sage wisdom and Eric Huff and Genevieve Graves for insight into issues in the SDSS photo pipeline. We thank Jessica Werk for insight into COS-Halos mass dependence and Molly Peeples for providing the data from \cite{Peeples:2014fe} for Figure \ref{molly}. 

JEGP was supported by HST-HF-51295.01A, provided by NASA through a Hubble Fellowship grant from STScI, which is operated by AURA under NASA contract NAS5-26555. BM is supported by NSF Grant AST-1109665 and work by LRC was supported by NASA Headquarters under the NASA Earth and Space Science Fellowship Program, grant NNX11AO09H.

Funding for the Sloan Digital Sky Survey (SDSS) and SDSS-II has been provided by the Alfred P. Sloan Foundation, the Participating Institutions, the National Science Foundation, the U.S. Department of Energy, the National Aeronautics and Space Administration, the Japanese Monbukagakusho, and the Max Planck Society, and the Higher Education Funding Council for England. The SDSS Web site is http://www.sdss.org/.

The SDSS is managed by the Astrophysical Research Consortium (ARC) for the Participating Institutions. The Participating Institutions are the American Museum of Natural History, Astrophysical Institute Potsdam, University of Basel, University of Cambridge, Case Western Reserve University, The University of Chicago, Drexel University, Fermilab, the Institute for Advanced Study, the Japan Participation Group, The Johns Hopkins University, the Joint Institute for Nuclear Astrophysics, the Kavli Institute for Particle Astrophysics and Cosmology, the Korean Scientist Group, the Chinese Academy of Sciences (LAMOST), Los Alamos National Laboratory, the Max-Planck-Institute for Astronomy (MPIA), the Max-Planck-Institute for Astrophysics (MPA), New Mexico State University, Ohio State University, University of Pittsburgh, University of Portsmouth, Princeton University, the United States Naval Observatory, and the University of Washington.

\bibliographystyle{yahapj}

\end{document}